# Accurately heat flow distribution based on a triple quantum dot

Yanchao Zhang[*]

*School of Science, Guangxi University of Science and Technology, Liuzhou 545006, People's Republic of China*

We theoretically propose a simple setup based on a three terminal triple quantum dot in the Coulomb blockade regime as a heat flow allocator to spatially separate heat flows along two different channels at arbitrary proportion. We show that a constant output heat flow ratio can be obtained in a wide range of system parameters and any ratio of the output heat flow, whether it is an integer ratio or a fractional ratio, can be obtained by directly adjusting the ratio of the energy-dependent tunneling rate.

[*] Email: zhangyanchao@gxust.edu.cn

Engineering and controlling heat flow at the nanoscale have attracted considerable interest in the fields of modern science and technology because of its fundamental and potential applications for the development of nanotechnologies [1-6]. In recent years, the double quantum-dot systems as new candidates show great promise in the fields of thermoelectric heat engines [7-11] and refrigerators [12-15]. Nowadays a significant effort has emerged that is devoted to designing a new generation of thermal functional devices [3, 16-19] base on the double quantum-dot systems in the Coulomb blockade regime. For example, thermal rectifiers/diodes [20, 21], thermal transistors [22], thermal logical gates [23]. In addition, related applications have been extended to the field of thermometry [24, 25] and quantum information [26, 27].

Recently, the research has been extended to a triple quantum dot system, which is motivated by the fact that a triple quantum dot system is important for quantum computation and further interesting theoretical predictions [28-33]. There have been theoretical and experimental reports on the triple quantum dot of serial or triangular geometry, which focus on charge rectification [34], the Aharonov-Bohm effect [35], charge frustration [36], and transport measurement [37-39]. Saraga and Loss first propose a solid-state entangler based on a triple quantum dot setup, that can spatially separate currents of spin-entangled electrons [28]. Vidan *et al*. presented an experimental realization of a triple quantum dot charging ratchet based on three tunnel-coupled quantum dots in the Coulomb blockade regime [34]. Recent experimental studies have shown that a triple quantum dot with triangular geometry can simultaneously measure electronic transport along two different paths [37] and the effect of interactions between the charge flowing through the two different paths have also been analyzed [38].

When consider three capacitively coupled quantum dots with triangular geometry in the Coulomb-blockade regime, the electron transport between the quantum dots is forbidden, but the heat transport is allowed by the Coulomb interaction [40]. In this respect, the triple quantum dots offer the possibility of analyzing new fascinating properties which are not presented in double quantum-dot systems. The latest research shows that this system can not only implement thermal diodes separately in two different paths but also perform more thermal management operations, such as heat flow swap, thermal switch, and heat path selector [40]. Recently, a thermal transistor based on this system has also been proposed [41].

In this Rapid Communication, we further propose that this triple quantum dot system can also be used as a solid-state heat flow allocator that can produce two arbitrarily proportional heat flows and allow them to be separated and extracted into two different channels for further processing.

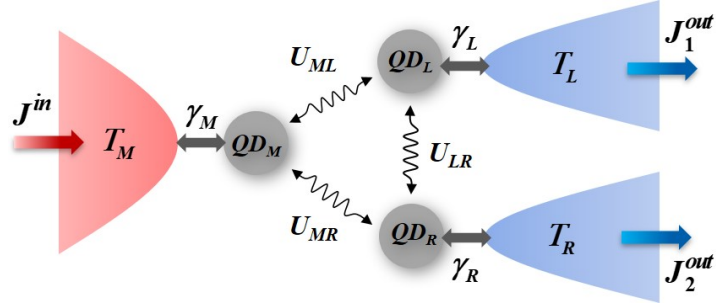

FIG. 1. (Color online) The model of heat flow allocator based on a triple quantum dot. Three quantum dots $QD_\alpha$ are coupled to three reservoirs at temperature $T_\alpha$ in the Coulomb blockade regime.

The model of heat flow allocator base on a triple quantum dot is illustrated in Fig. 1, where three quantum dots are placed in the center of the device, such that they are Coulomb-coupled to each other and interact only through the long-range Coulomb force such that they can only exchange energy but no particles. Each of quantum dots $\alpha$ ($\alpha = M, L, R$) is connected to its reservoir with temperature $T_\alpha$ via a tunnel barrier $\gamma_\alpha$ and permits particle and energy exchanges. The electron transport between the quantum dots is forbidden but the Coulomb interaction $U_{\alpha\beta}$ ($\alpha, \beta = M, L, R$  $\alpha \neq \beta$) between the quantum dots $\alpha$ and $\beta$ allows through electron tunneling into and out of quantum dots to transmit the heat between the reservoirs with a temperature difference. Such a triple quantum dot configuration constitutes two different heat flow channels separating the input heat flow $J^{in}$ into two output heat flows $J_1^{out}$ and $J_2^{out}$. Channel 1 transmits heat $J_1^{out}$ from the reservoir $M$ to the reservoir $L$ by the Coulomb interaction $U_{ML}$ between the quantum dots $M$ ($QD_M$) and $QD_L$, and Channel 2 transmits heat $J_2^{out}$ from the reservoir $M$ to $R$ by the Coulomb interaction $U_{MR}$ between the $QD_M$ and $QD_R$, this creates two heat flows that are spatially separated. The aim of the heat flow allocator is to extract input heat flow from the $QD_M$, by Coulomb interaction into the neighboring dot $QD_L$ and the other one $QD_R$, and finally transport them into the two output heat flows with arbitrary proportions.

The triple quantum dot system is described by the total Hamiltonian $H = H_D + H_R + H_T$, where $H_D$, $H_R$, and $H_T$ are the Hamiltonians of the quantum dot, the reservoir, and the tunneling Hamiltonian between quantum dot and the reservoir, respectively. The Hamiltonian Coulomb-coupled quantum dots is described by

$$H_D = \sum_{\alpha=M,L,R} \varepsilon_\alpha d_\alpha^\dagger d_\alpha + \sum_{\alpha,\beta=M,L,R\ \alpha\neq\beta} U_{\alpha\beta} d_\alpha^\dagger d_\alpha d_\beta^\dagger d_\beta, \tag{1}$$

where $d^\dagger(d)$ is the creation (annihilation) operator of quantum dot with a single energy level $\varepsilon_\alpha$. The Hamiltonian of reservoir is defined as

$$H_R = \sum_k \sum_{\alpha=M,L,R} \varepsilon_{\alpha k} c_{\alpha k}^\dagger c_{\alpha k}, \tag{2}$$

where $\varepsilon_{\alpha k}$ is the energy of the noninteracting reservoir electrons with continuous wave number $k$, $c^\dagger(c)$ denote the creation (annihilation) operators of heat reservoir. Finally, the tunneling Hamiltonian between the quantum dot and the reservoir is given by

$$H_T = \sum_k \sum_{\alpha=M,L,R} \left( t_{\alpha k} c_{\alpha k}^\dagger d_\alpha + t_{\alpha k}^* d_\alpha^\dagger c_{\alpha k} \right), \tag{3}$$

where $t_{\alpha k}$ and its conjugate $t_{\alpha k}^*$ denote the tunneling amplitudes.

The triple quantum dot system is denoted by the charge configuration $|n_M, n_L, n_R\rangle$, where $n_\alpha$ are the occupation number of quantum dots $\alpha$. In the Coulomb blockade regime, each of these quantum dots can be occupied only by zero or one electron ($n_\alpha = 0,1$). Thus, the dynamics of the quantum dot system is characterized by eight charge states labeled as $|1\rangle = |0,0,0\rangle$, $|2\rangle = |1,0,0\rangle$, $|3\rangle = |0,1,0\rangle$, $|4\rangle = |0,0,1\rangle$, $|5\rangle = |1,1,0\rangle$, $|6\rangle = |1,0,1\rangle$, $|7\rangle = |0,1,1\rangle$, and $|8\rangle = |1,1,1\rangle$. The occupation probabilities for eight charge states are given by the diagonal elements of the density matrix, $\boldsymbol{\rho} = (\rho_1, \rho_2, \rho_3, \rho_4, \rho_5, \rho_6, \rho_7, \rho_8)^T$. In the limit of weak tunneling coupling ($\hbar\gamma \ll k_B T$), the broadening of energy levels can be neglected and the transmission through tunnel barriers is well described by sequential tunneling rates. The off-diagonal density matrix elements do not contribute to steady state transport and can be neglected. Thus, the time evolution of occupation probabilities is given by a master equation. The matrix form can be written as $d\rho/dt = M\rho$, where $M$ denotes the matrix containing the transition rates and is given by Fermi's golden rule. The steady-state heat flows from electronic reservoir $\alpha$ to quantum dot $\alpha$ are given by the stationary solution of the master equation $M\bar{\rho} = 0$. Because the middle reservoir drives the transitions of charge states including $|1\rangle \leftrightarrow |2\rangle$, $|3\rangle \leftrightarrow |5\rangle$, $|4\rangle \leftrightarrow |6\rangle$ and $|7\rangle \leftrightarrow |8\rangle$, and the input heat flow $J^{in}$ is expressed as

$$J^{in} = (\varepsilon_M - \mu_M)(\Gamma_{21}\bar{\rho}_1 - \Gamma_{12}\bar{\rho}_2) + (\varepsilon_M + U_{MR} - \mu_M)(\Gamma_{53}\bar{\rho}_3 - \Gamma_{35}\bar{\rho}_5) \\ + (\varepsilon_M + U_{ML} - \mu_M)(\Gamma_{64}\bar{\rho}_4 - \Gamma_{46}\bar{\rho}_6) + (\varepsilon_M + U_{MR} + U_{ML} - \mu_M)(\Gamma_{87}\bar{\rho}_7 - \Gamma_{78}\bar{\rho}_8). \quad (4)$$

The left reservoir induces the transitions of charge states including $|1\rangle \leftrightarrow |3\rangle$, $|2\rangle \leftrightarrow |5\rangle$, $|4\rangle \leftrightarrow |7\rangle$ and $|6\rangle \leftrightarrow |8\rangle$, the output heat flow $J_1^{out}$ through channel 1 is given by

$$J_1^{out} = (\varepsilon_L - \mu_L)(\Gamma_{31}\bar{\rho}_1 - \Gamma_{13}\bar{\rho}_3) + (\varepsilon_L + U_{ML} - \mu_L)(\Gamma_{52}\bar{\rho}_2 - \Gamma_{25}\bar{\rho}_5) \\ + (\varepsilon_L + U_{LR} - \mu_L)(\Gamma_{74}\bar{\rho}_4 - \Gamma_{47}\bar{\rho}_7) + (\varepsilon_L + U_{ML} + U_{LR} - \mu_L)(\Gamma_{86}\bar{\rho}_6 - \Gamma_{68}\bar{\rho}_8). \quad (5)$$

The right reservoir triggers the transitions of charge states including $|1\rangle \leftrightarrow |4\rangle$, $|2\rangle \leftrightarrow |6\rangle$, $|3\rangle \leftrightarrow |7\rangle$ and $|5\rangle \leftrightarrow |8\rangle$, thus the output heat flow $J_2^{out}$ through channel 2 is written as

$$J_2^{out} = (\varepsilon_R - \mu_R)(\Gamma_{41}\bar{\rho}_1 - \Gamma_{14}\bar{\rho}_4) + (\varepsilon_R + U_{MR} - \mu_R)(\Gamma_{62}\bar{\rho}_2 - \Gamma_{26}\bar{\rho}_6) \\ + (\varepsilon_R + U_{LR} - \mu_R)(\Gamma_{73}\bar{\rho}_3 - \Gamma_{37}\bar{\rho}_7) + (\varepsilon_R + U_{MR} + U_{LR} - \mu_R)(\Gamma_{85}\bar{\rho}_5 - \Gamma_{58}\bar{\rho}_8). \quad (6)$$

In Eqs. (4)-(6), $\Gamma_{ji} = \gamma_\alpha f_\alpha(E_{ji})$ ($i, j = 1, 2, \cdots, 8$ and $i < j$) is the transition rate from the charge state $|i\rangle$ to $|j\rangle$ with an electron from the electronic reservoir $\alpha$ into the quantum dot $\alpha$, and $\Gamma_{ij} = \gamma_\alpha (1 - f_\alpha(E_{ji}))$ when an electron leaves the quantum dot $\alpha$ into the electronic reservoir $\alpha$, where $f_\alpha(x) = \{\exp[(x - \mu_\alpha)/(k_B T_\alpha)] + 1\}^{-1}$ is the Fermi-Dirac distribution function with the chemical potential $\mu_\alpha$ and temperature $T_\alpha$, $E_{ji} = E_j - E_i$, $E_i$ is the energy of the charge state $i$, $k_B$ is the Boltzmann constant, and $\gamma_\alpha$ is the energy-dependent tunneling rates between the quantum dots and the respective electronic reservoirs. In the present model, the electron transport between the quantum dots is forbidden. Thus, the heat flows fulfill $J^{in} + J_1^{out} + J_2^{out} = 0$, which complies with the energy conservation.

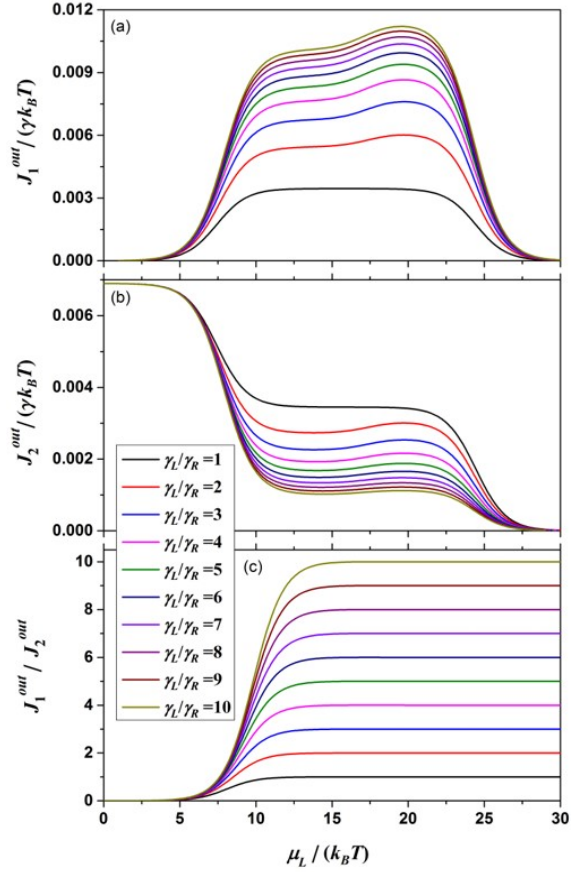

FIG. 2. (Color online) The output heat flows $J_1^{out}$, $J_2^{out}$ and heat flow ratio $J_1^{out}/J_2^{out}$ as a function of the chemical potential $\mu_L$. The parameters: $\gamma_M = \gamma_R$, $\varepsilon_\alpha = 0$, $\mu_M = 0$, $\mu_R = 15k_BT$, $U_{\alpha\beta} = 25k_BT$, $T_M = 500\text{mK}$, and $T_L = T_R \equiv T = 150\text{mK}$.

In Fig. 2 we plot the output heat flows $J_1^{out}$, $J_2^{out}$ and heat flow ratio $J_1^{out}/J_2^{out}$ versus the chemical potential $\mu_L$ for different ratio of the energy-dependent tunneling rate $\gamma_L/\gamma_R$. It clearly shown that the $J_1^{out}$ and $J_2^{out}$ are obviously dependent on the chemical potential. However, when the chemical potential $\mu_L > 15k_BT$, we can note that over a wide range of system parameters, the heat flow ratio only depends on the ratio of the energy-dependent tunneling rate $\gamma_L/\gamma_R$, which means that we can get arbitrarily proportional output heat flow by adjusting the ratio of the energy-dependent tunneling rate $\gamma_L/\gamma_R$.

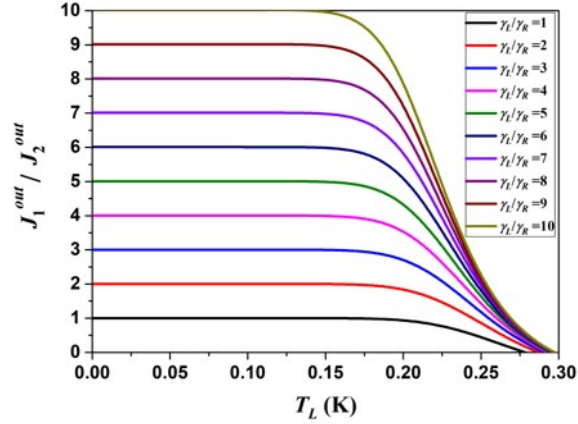

FIG. 3. (Color online) The heat flow ratio $J_1^{out}/J_2^{out}$ as a function of the temperature $T_L$. The parameters: $\gamma_M = \gamma_R$, $\varepsilon_\alpha = 0$, $\mu_M = 0$, $\mu_L = \mu_R = 15k_BT$, $U_{\alpha\beta} = 25k_BT$, $T_M = 500\text{mK}$, and $T_R \equiv T = 150\text{mK}$.

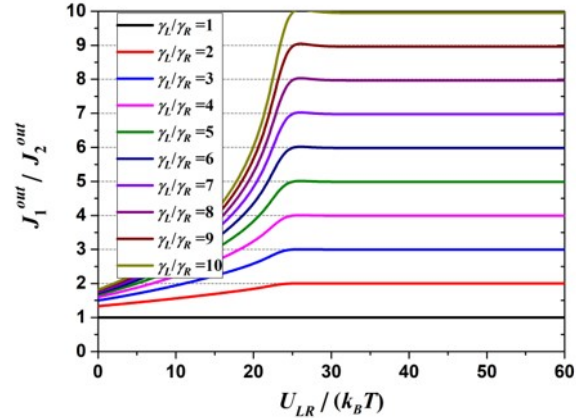

FIG. 4. (Color online) The heat flow ratio $J_1^{out}/J_2^{out}$ as a function of the Coulomb interaction $U_{LR}$. The parameters: $\gamma_M = \gamma_R$, $\varepsilon_\alpha = 0$, $\mu_M = 0$, $\mu_L = \mu_R = 15k_BT$, $U_{LM} = U_{RM} = 25k_BT$, $T_M = 500\text{mK}$, and $T_L = T_R \equiv T = 150\text{mK}$.

In Figs. 3 and 4, we show that effect on the heat flow ratio $J_1^{out}/J_2^{out}$ by the temperature $T_L$

and the Coulomb interaction $U_{LR}$, respectively. In order to get a constant proportion of the output heat flow, the temperature should be $T_L < 150\text{mK}$ or the Coulomb interaction $U_{LR} > 30k_B T$. We can note that over a wide range of system parameters, the heat flow ratio remains constant as long as $T_L < 150\text{mK}$ and $U_{LR} > 30k_B T$. In these cases, the heat flow ratio $J_1^{out}/J_2^{out}$ is independent of the temperature $T_L$ and the Coulomb interaction $U_{LR}$, but remarkably depends on the ratio of the energy-dependent tunneling rate $\gamma_L/\gamma_R$ and equals to the ratio of the energy-dependent tunneling rate $\gamma_L/\gamma_R$. Thus, the heat flow allocator proposed here does not require additional regulatory external parameters to achieve a constant proportion of the output heat flow once these conditions are fulfilled.

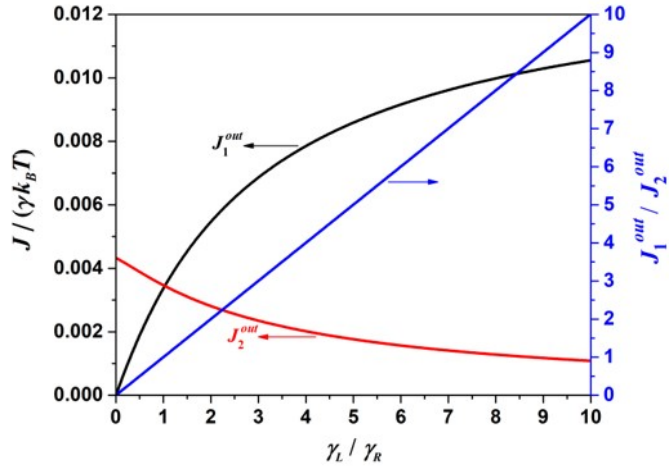

FIG. 5. (Color online) The output heat flows $J_1^{out}$, $J_2^{out}$ and the heat flow ratio $J_1^{out}/J_2^{out}$ as a function of the ratio of the energy-dependent tunneling rate $\gamma_L/\gamma_R$. The parameters: $\gamma_M = \gamma_R$, $\varepsilon_\alpha = 0$, $\mu_M = 0$, $\mu_L = 25k_B T$, $\mu_R = 15k_B T$, $U_{LM} = U_{RM} = 25k_B T$, $U_{LR} = 35k_B T$, $T_M = 500\text{mK}$, $T_L = 100\text{mK}$ and $T_R \equiv T = 150\text{mK}$.

The output heat flows $J_1^{out}$, $J_2^{out}$ and the heat flow ratio $J_1^{out}/J_2^{out}$ as a function of the ratio of the energy-dependent tunneling rate $\gamma_L/\gamma_R$ are shown in Fig. 5 for $\mu_L = 25k_B T$, $U_{LR}=35k_{B}T$, and $T_{L}=100\text{mK}$. It is found that the output heat flow $J_1^{out}$ increases with the increase of the $\gamma_L/\gamma_R$, while the output heat flow $J_2^{out}$ decreases with the increase of the $\gamma_L/\gamma_R$. In addition, when $\gamma_L/\gamma_R<1$, the output heat flow $J_1^{out}<J_2^{out}$, this indicates the heat flow ratio $J_1^{out}/J_2^{out}<1$, however, in the region of $\gamma_L/\gamma_R>1$, $J_1^{out}>J_2^{out}$ and $J_1^{out}/J_2^{out}>1$. In particular, the output heat flow $J_1^{out}=J_2^{out}$ when $\gamma_L/\gamma_R=1$. Remarkably, over the entire interval, the heat flow ratio is equal to the ratio of the energy-dependent tunneling rate, i.e., $J_1^{out}/J_2^{out}=\gamma_L/\gamma_R$. This means that any ratio of the output heat flow can be obtained by directly adjusting the ratio of the energy-dependent tunneling rate, whether it is an integer ratio or a fractional ratio. This is a nice property for practical applications, since it does not require additional regulatory external parameters.

In summary, we proposed a heat flow allocator that can produce two arbitrarily proportional heat flows and allow them to be separated and extracted into two different channels. We demonstrated that the device can achieves any ratio of the output heat flow by directly adjusting the ratio of the energy-dependent tunneling rate over a wide range of system parameters. These results should have important implications in providing the design principle for solid-state heat flow allocator devices and may open up potential applications for the heat flow control at the nanoscale.

Acknowledgments
This work was supported by the National Natural Science Foundation of China (No. 11947010), the Science and Technology Base and Talent Project of Guangxi (No. AD19110104), and the Guangxi University of Science and Technology Foundation for PhDs (No. 18Z11).